# Intermediate Bands in Zero-Dimensional Antimony Halide Perovskites

David A. Egger[*]

*Institute of Theoretical Physics, University of Regensburg, 93040 Regensburg, Germany*

\* Correspondence: david.egger@physik.uni-regensburg.de

**Abstract**

Using density functional theory, the structural and electronic-structure properties of a recently discovered, zero-dimensional antimony halide perovskite are studied. It is found that the herein considered material EtPySbBr$_6$ exhibits very promising electronic-structure properties: a direct band gap close to the peak of the solar spectrum and effective masses allowing for efficient carrier transport of electrons in particular. These results are rationalized by analysis of the electronic structure, which reveals the formation of intermediate bands due to orbital-hybridization effects of the Sb *s*-states. This study shows that the formation of intermediate bands can lead to highly favorable electronic-structure properties of zero-dimensional perovskites and discusses the possibility of fabricating lead-free HaPs with promising optoelectronic properties by targeted substitution of ions and emergence of intermediate bands. These insights are important when understanding and further enhancing the capabilities of antimony and other promising lead-free compounds.

**TOC Figure**

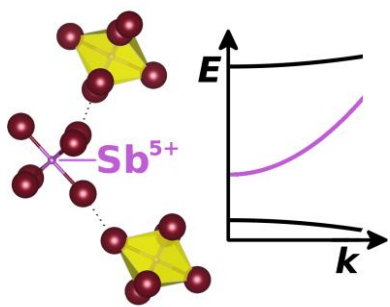

Three-dimensional (3-D) halide perovskites (HaPs) are crystalline semiconductors with $A^{1+}B^{2+}X^{-1}{}_3$ stoichiometry in which $X$ is a halide ion, $A$ is typically a large organic or inorganic cation, and $B$ a divalent metal anion.[1] The ideal 3-D HaP structure implies geometrical constraints for the radii of ions and formation of a network of corner-sharing $BX_6$ octahedra with voids occupied by $A$-site cations.[2] Within this structure, the ions exhibit large coordination numbers in general, and strong covalency among the $B$ and $X$ ions in particular, which can give rise to preferred optoelectronic properties: the prototypical MAPbI$_3$ (where MA stands for methylammonium) exhibits a direct band gap with strong optical absorption close to the peak of the solar spectrum as well as small effective masses of both electrons and holes.[3–7] While several of the intriguing properties of HaPs are currently not fully understood,[8] it is clear that their outstanding electronic-structure characteristics are central to their successes as materials for high-efficiency optoelectronic devices including photovoltaic (PV) cells.

One major obstacle between HaPs and their commercialization as materials for large-scale energy devices is that their most efficient variants include the poisonous element lead. Many promising routes towards Pb-free HaPs have been explored in recent years,[9] including inverse[10] and double perovskites[11–15] as well as replacing Pb by elements such as tin,[16–21] bismuth,[22–25] or antimony.[26–29] However, the geometrical constraints imposed by the necessity of forming an ideal perovskite lattice, as well as the common oxidation states of these alternative cations (*e.g.*, +3 of Bi and Sb), still render difficult the discovery of lead-free materials that could rival the optoelectronic properties of prototypes such as MAPbI$_3$. This is true in particular for the aforementioned beneficial electronic-structure features of 3-D lead-based HaPs, *i.e.*, a direct band gap that is close to optimal for solar light absorption and dispersive electronic bands allowing for efficient carrier transport.

An alternative route towards novel halide compounds beyond ideal 3-D HaPs is reducing the inter-octahedral connectivity in the perovskite lattice and fabrication of lower-dimensional structures.[30] In these crystals, the octahedra of the former perovskite are isolated and inter-octahedral conjugation is strongly reduced along specific crystalline directions where the lattice deviates from the ideal 3-D HaP structure. The extreme case is a crystal exhibiting fully isolated $BX_6$ octahedra and minimal inter-octahedral covalency,[31] which were referred to as *zero-dimensional (0-D) HaPs* in the literature.[30–32] Because the conjugation among the octahedra in 0-D HaPs is small, their reported band gaps were found to be much larger and band dispersions to be much smaller than in 3-D HaPs,[31,33,34] which has thus far drastically limited their use in efficient optoelectronic devices.

In this letter, a recently discovered[35] lead-free variant of 0-D HaPs that uses antimony not in its previously considered +3 oxidation state but in the +5 oxidation state as a *B*-site cation, and has shown enormous potential as a PV material, is studied. Specifically, the structural and electronic properties of a compound with $A^{1+}B^{5+}X^{-1}_6$ stoichiometry involving ethylpyridinium (EtPy) and $Sb^{5+}$ cations as well as Br anions,[35] is investigated using density functional theory (DFT). The resulting $EtPySbBr_6$ crystal (see Fig. 1) exhibits very promising electronic-structure properties: a direct band gap close to the peak of the solar spectrum and effective masses that allow for efficient carrier transport of especially electrons. These results are explained by formation of intermediate bands in $EtPySbBr_6$ due to orbital-hybridization effects of the Sb *s*-states, which demonstrates the important role of the $Sb^{+5}$ ion for the outstanding electronic-structure properties of $EtPySbBr_6$. The discussion of these insights suggests promising routes to material fabrication by ionic substitution and formation of intermediate bands in the quest of discovering lead-free HaP compounds.

Periodic DFT calculations were performed using the VASP planewave code[36] and the Perdew-Burke-Ernzerhof (PBE) functional,[37] augmented by pair-wise dispersion interactions of the Tkatchenko-Scheffler scheme.[38] Core-valence electron interactions were treated within the projector-augmented wave (PAW) formalism[39] using the program supplied "normal" version of the PAW potentials. Optimizations of the lattice constants and single-point total energy calculations were performed with a planewave cutoff energy of 700 eV and 400 eV, respectively. A 1x4x1 Γ-centered *k*-point grid and convergence criteria of $10^{-6}$ eV for the total energy and $10^{-2}$ eV/Å for residual forces and stresses were employed in all calculations. For determining cell parameters and atomic coordinates of $EtPySbBr_6$, the experimentally reported crystal structure[35] was used as a starting point in optimizations with the GADGET tool using internal coordinates.[40] The experimentally determined ratio of the unit-cell vectors was imposed in these calculations, a constraint that was tested to have virtually no effect on the optimized unit-cell volume. It is noted that the unit cell of EtPySbBr$_6$ contains 200 atoms, which renders it computationally challenging. To nevertheless provide an accurate estimation of its electronic band-structure and density of states (DOS), the screened hybrid functional of Heyd, Scuseria and Ernzerhof (HSE)[41,42] was used since it can partially correct the underestimation of DFT-calculated band gaps and improve descriptions of differently delocalized electronic states. For visualizations of the structure and analysis of the electronic properties the VESTA[43] and sumo[44] programs were applied, respectively. Using the latter, a parabolic fit of the electronic band structure at the valence band maximum and conduction band minimum has been applied to determine the effective masses.

The EtPySbBr$_6$ crystal is monoclinic and belongs to the P2$_1$/c space group, as reported in ref. 35. DFT-optimized structural parameters of EtPySbBr$_6$ (see Fig. 1), *i.e.*, the length of the unit-cell vectors, are $a = 20.92$ Å, $b = 7.41$ Å, $c = 21.93$ Å, which are in very good agreement with the single-crystal structure reported from diffraction experiments at room temperature ($a = 20.72$ Å, $b = 7.34$ Å, $c = 21.72$ Å). The effect of dispersive interactions in determining structural parameters is found to be large, as structural optimization performed using the PBE functional without adding dispersive corrections yields much larger unit cells ($a = 21.81$ Å, $b = 7.73$ Å, $c = 22.86$ Å), similar to the case of lead-based HaPs.[45–49] This finding is a strong indication that the EtPySbBr$_6$ lattice is at least partially held together by weak dispersive interactions including contributions from van-der-Waals and H-bonding.

We continue by analyzing the fully-relaxed structure of EtPySbBr$_6$ (see Fig. 1) in more detail. From a structural perspective, it is interesting to compare its crystal structure to ideal 3-D HaPs: EtPySbBr$_6$ is similar to the ideal 3-D HaP structure because a network of highly-coordinated ions, involving inorganic octahedra and space-filling *A*-site cations, is formed. However, it is clearly *not identical to it and does not reflect a perovskite structure*, because the octahedra are not corner-sharing and the [SbBr$_6$]$^{-1}$ units appear to be isolated. Because this is the case also in common Pb-based 0-D HaPs, the EtPySbBr$_6$ crystal may be described best as reflecting a 0-D HaP. However, in contrast to the latter EtPySbBr$_6$ was found to show favorable optoelectronic properties, which could be related to non-negligible inter-octahedral interactions. In line with this reasoning and following ref. 35, it is also found here that the formation of a network of covalently-linked octahedra appears to be the case in EtPySbBr$_6$: First, the optimized Sb-Br distances (2.59–2.63 Å) show that intra-octahedral covalent bonds are formed. More importantly, the Br-Br bonds lengths between neighboring octahedra are calculated to be 3.26–3.51 Å, which demonstrates significant inter-octahedral covalency in EtPySbBr$_6$, as schematically shown in the octahedral pattern (see Fig. 2). While the Br-Br bond lengths are spatially isotropic to a large degree, a pattern of alternating shorter and longer bonds is also found, as shown in the supporting information (SI).

To understand better the potential octahedral covalency in EtPySbBr$_6$ as well as the microscopic origin of its favorable optoelectronic properties, we first consider the electronic band structure shown in Fig. 3. The valence bands are closely spaced over a range of ~2.3 eV, and the conduction bands show two groups of bands separated by an energy gap of ~1.8 eV, *i.e.*, eight dispersive bands and eight higher-lying flat ones. This band structure resembles what is known as an "intermediate band-gap material": such compounds are characterized by *valence, conduction and intermediate bands* allowing for distinct optical excitations that can in principle lead to power conversion efficiencies of PV devices beyond the Shockley-Queisser limit.[50] It is noted that realizing the latter in practice has been challenging, and for the specific case of Sn-based 0-D HaPs may require the discovery of alternative materials with an even smaller optical gap.[50] Most importantly, *a direct gap between the valence and intermediate band* is found in EtPySbBr$_6$ at the center of the Brillouin zone, amounting to 1.5 eV in the HSE calculations, in close agreement with the 1.65 eV optical gap reported experimentally.

Following this analysis of the band structure it would be tempting to discuss its relation to the optical absorption profile of EtPySbBr$_6$ as reported in ref. 35. The optical spectrum showed several distinct features across a wide energy range of 1.6 – 3.9 eV. While this is in agreement with the expectation born from the band structure of EtPySbBr$_6$ and the formation of intermediate bands, it is to be noted that the electronic-structure was calculated using ground-state DFT. Indeed, a full theoretical characterization of the excited-state properties of the material would be required to provide further insight. In this context, it would be interesting not only to provide a theoretical estimate on the absorption strength compared to other HaPs, but also to understand the role of excitonic effects. Such calculations go beyond the scope of the present study considering the large system size of EtPySbBr$_6$ and further conceptual issues, such as the possibility that nuclear vibrations may provide screening in the excited state.

These results demonstrate that in contrast to the case of 0-D Pb-based HaPs such as Cs$_4$PbBr$_6$,[33] efficient light absorption close to the peak of the solar spectrum is possible in EtPySbBr$_6$. Furthermore, it is interesting to compare this finding to known results[7,17] for 3-D Pb-based HaPs, as the band gap of EtPySbBr$_6$ *is substantially lower* than that of MAPbBr$_3$ and in fact reasonably close to the one of MAPbI$_3$. Thus, while reducing the dimensionality of Pb-based HaPs typically *resulted in larger band gaps* due to reduced orbital overlap,[31,33,34] the 0-D antimony HaP EtPySbBr$_6$ surprisingly *exhibits a smaller band gap* compared to the 3-D Pb-based HaP bromide variants. The formation of intermediate bands can explain this result, as by its very nature it implies band gap reduction. The electronic structure and orbital hybridization in 0-D antimony HaPs therefore appear to be strongly influenced by the formation of intermediate bands, which is fundamentally different to the case of the Pb-based HaP analogues.

To better understand the formation of the intermediate bands, we examine the orbital hybridization in EtPySbBr$_6$ and consider the atomic-orbital projected density of states (PDOS) shown in Fig. 3b. The upper-lying valence bands are dominated by Br *p*-states that hybridize only weakly with other orbitals. The intermediate bands, however, involve much stronger hybridization, especially between Br *p*- and Sb *s*-states. Note that the EtPy molecular orbitals hybridize weakly with the Br *p*-states in the lower-lying valence bands and make up most of the higher-lying flatter conduction bands. For the PDOS of all relevant atomic orbitals, see the SI.

It is now possible to rationalize the intriguing electronic-structure features of EtPySbBr$_6$. The occupied states of the Br atoms show rather weak mutual hybridization, as can be seen in real-space representations of the charge density associated with the valence band, shown in Fig. 3c, left panel. Consequently, the valence band dispersion is rather weak and the hole effective mass, $m_h^*$, is rather large ($m_h^* \sim 1.1 m_e$ where $m_e$ is the electron mass). In contrast, the low-lying unoccupied *s*-states of Sb hybridize with the unoccupied Br *p*-states (see real space representation in Fig. 3c, right panel) resulting in the formation of intermediate bands. Consequently, the latter exhibit appreciable band dispersion and low electron effective masses, $m_e^* \sim 0.5 m_e$. Furthermore, along the direction of lower dispersion of the intermediate band, *i.e.*, from $R_2$ to $T_2$ in Fig. 3a, the orbital hybridization among neighboring octahedra is reduced (see SI). Therefore, the *intrinsic presence of the low-lying unoccupied Sb s-states and its hybridization with anionic states are important*: they result in the intermediate band-gap character of EtPySbBr$_6$ as well as its very promising optoelectronic properties (direct band gap of ~1.5 eV and low $m_e^*$), similar to what has been reported for oxide perovskites and doping by Bi$^{5+}$ ions.[51]

From these results, the implications of intermediate band formation in HaPs may finally be discussed. It has been shown that (partially) lifting the geometrical constraints implicit to the ideal 3-D HaP structure reduces the orbital overlap and hybridization of ions in the perovskite lattice.[31,33,34] This leads to less preferred optoelectronic properties of 0-D HaPs compared to their 3-D analogues. Furthermore, using alternative ions in lead-free systems has proven to be very challenging. Here, it is suggested that these obstacles can be circumvented by the use of Sb$^{5+}$ ions with their *intrinsically low-lying s-states to introduce an intermediate band in the electronic structure*. The result of such an approach is the 0-D antimony HaP crystal EtPySbBr$_6$ with highly promising electronic-structure properties. 0-D antimony HaPs of this kind may just be one example of a potentially wide range of new compounds with ions that would result in intermediate bands as part of the perovskite lattice. These could be fabricated without the need of being fully restricted to the ideal 3-D HaP structure, as the formation of intermediate bands may still result in favorable optoelectronic properties. Whether these will be sufficiently beneficial for use in highly efficient PV devices in general, and whether this is the case for 0-D Sb HaP in particular, will also depend on other material properties. These include the strength of optical absorption, the presence of defects and the role of non-radiative recombination, which need to be addressed in future studies.

In summary, the structural and electronic-structure properties of a highly-promising lead-free 0-D HaP crystal that is based on antimony ions were studied using theoretical calculations based on DFT. Most importantly, in contrast to 0-D Pb-based HaPs the here-studied EtPySbBr$_6$ crystal was found to exhibit very promising electronic-structure properties: a direct band gap close to the peak of the solar spectrum and effective masses allowing for efficient carrier transport of electrons, which are key features for efficient PV energy conversion. These findings were rationalized by the formation of intermediate bands in EtPySbBr$_6$ as a consequence of the presence and orbital-hybridization effects of the Sb *s*-states. Thus, using ions that introduce intermediate bands emerges as a route to fabricate 0-D and other lower-dimensional lead-free HaPs with promising optoelectronic properties. In addition, these findings highlight the potential, and support the understanding, of prospects and challenges of Sb-based HaPs as a platform for efficient and sustainable optoelectronic materials.

## Acknowledgements

I would like to thank Prof. Andrew Rappe (University of Pennsylvania) for illuminating discussions. Funding provided by the Alexander von Humboldt Foundation in the framework of the Sofja Kovalevskaja Award endowed by the German Federal Ministry of Education and Research is acknowledged. Computing time granted by the John von Neumann Institute for Computing (NIC) and provided on the supercomputer JURECA[52] at Jülich Supercomputing Centre (JSC) is appreciated as well.

**Figure 1**

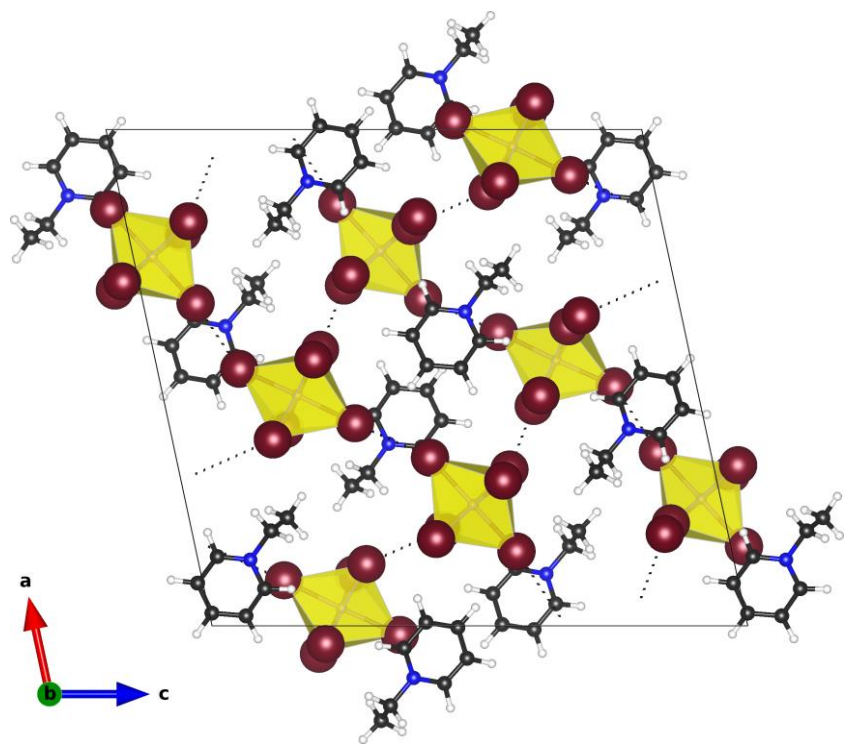

Fig. 1: Schematic structural representation of the zero-dimensional antimony halide perovskite EtPySbBr$_6$ consisting of Br (dark red), carbon (black), nitrogen (dark blue), hydrogen (white) and Sb atoms (magenta). The latter are inside yellow-shaded octahedra and atoms belonging to more than a unit cell are shown for easy visualization. Dashed black lines indicate connectivity between Br atoms of neighboring octahedra and thin black lines describe the boundary of the unit cell.

**Figure 2**

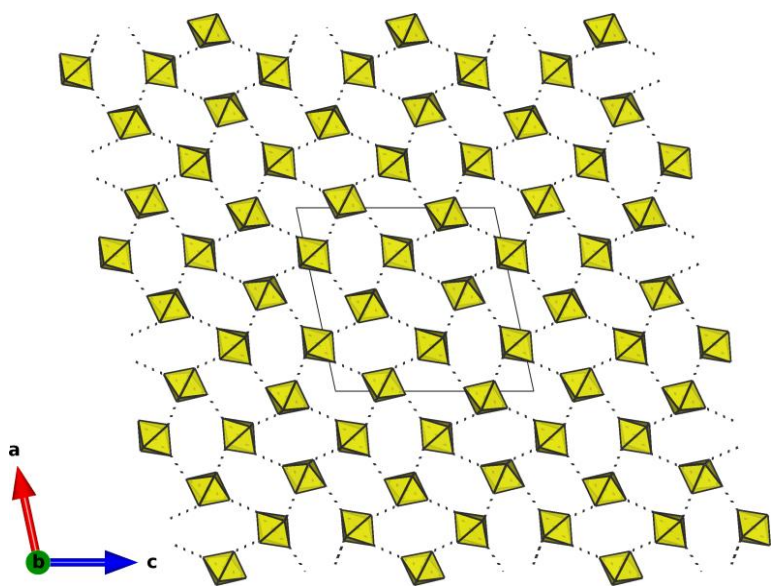

Fig. 2: Schematic structural representation of the pattern of octahedra (shown in yellow) across several unit cells of EtPySbBr$_6$. The EtPy molecular units are omitted, dashed black lines indicate connectivity between Br atoms of neighboring octahedra and thin black lines describe the boundary of the unit cell.

**Figure 3**

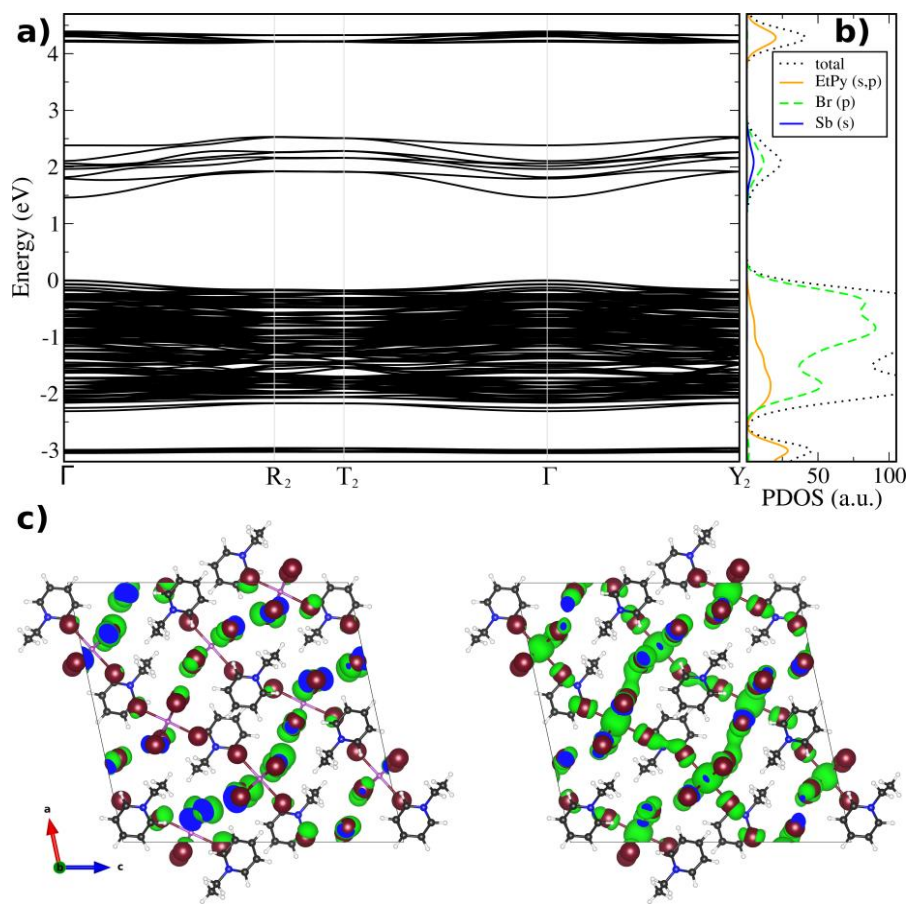

Fig. 3: a) DFT-calculated electronic band structure of EtPySbBr$_6$ along high-symmetry directions of the Brillouin zone. The energy axis has been aligned to the valence band maximum in the band-structure; Γ, R$_2$, T$_2$ and Y$_2$ denote (0,0,0), (-0.5,-0.5,0.5), (0,-0.5,0.5), and (0,-0.5,0), respectively. b) Total density of states of EtPySbBr$_6$ (black dotted line) and density of states projected onto atomic orbitals (PDOS) of EtPy (*s*- and *p*-states, orange line), Sb (*s*-states, blue line) and Br (*p*-states, green dashed line) showing Sb-Br hybridization and formation of intermediate bands. c) Real-space representation of the charge density associated with the valence band (left) and intermediate band (right) of EtPySbBr$_6$, showing weaker Br-Br and stronger Sb-Br hybridization, respectively.